# Attention-Based Recommendation On Graphs

Taher Hekmatfar, Saman Haratizadeh, Parsa Razban, Sama Goliaei
{taher.hekmatfar, haratizadeh, parsa.razban, sgoliaei}@ut.ac.ir

**ABSTRACT**
Graph Neural Networks (GNN) have shown remarkable performance in different tasks. However, there are a few studies about GNN on recommender systems. GCN as a type of GNNs can extract high-quality embeddings for different entities in a graph. In a collaborative filtering task, the core problem is to find out how informative an entity would be for predicting the future behavior of a target user. Using an attention mechanism, we can enable GCNs to do such an analysis when the underlying data is modeled as a graph. In this study, we proposed GARec as a model-based recommender system that applies an attention mechanism along with a spatial GCN on a recommender graph to extract embeddings for users and items. The attention mechanism tells GCN how much a related user or item should affect the final representation of the target entity. We compared the performance of GARec against some baseline algorithms in terms of RMSE. The presented method outperforms existing model-based, non-graph neural networks and graph neural networks in different MovieLens datasets.

## 1.     Introduction

Graphical representations are being used as a powerful approach to modeling related entities in a variety of applications, including friendship analysis in social media (Silva et al., 2010), modeling protein-protein interactions in biology (Hamilton, Ying, & Leskovec, 2017), constructing knowledge graphs for reasoning (Ji et al., 2021), and recommending items to users in e-commerce domain (Shams & Haratizadeh, 2017). In the recommender systems domain, graph-based modeling of data can also help to overcome the sparsity problem by expanding the concept of similarity to cover indirect relations among entities as well (Shams & Haratizadeh, 2017). A variety of methods exist for defining similarity measures in recommender system graphs, including meta-paths analysis (Shi et al., 2014; Xie et al., 2015) or random walk simulation and PageRank calculation (J. Chen et al., 2019; Musto et al., 2017; Palumbo et al., 2018; Shams & Haratizadeh, 2017). However, the main challenge in these methods is that it is not clear how to recognize reliable and unreliable paths in the graph that can reflect the true similarity among entities. (Shams & Haratizadeh, 2018).

Recently there has been an increasing interest in applying Deep Learning based methods on GRSs to replace the direct similarity calculation process with a model-based approach that embeds the user/item nodes and extracts recommendations using those embeddings. Graph Neural Network (GNN) is a general approach to extract embeddings for nodes in a graph that could be applied on any graphs, including user and item nodes in recommender graphs (Ying, You, et al., 2018a). These methods use the structural information to capture direct and indirect relationships among entities and combine it with local node information to extract informative embeddings (Fan et al., 2019)

This paper proposes a novel Graph Neural Network to embed users and item nodes in a heterogeneous user-item graph. This method benefits an attention mechanism whose effectiveness in extracting embedding has been shown in various domains like machine translation (Vaswani et al., 2017), text classification (Yao, Mao, & Luo, 2019), and protein networks (Veličković et al., 2017). The main idea here is to recognize and use the more informative relevant nodes while extracting embedding for a target node. The proposed method has the advantage of both similarity-based and model-based approaches. Like a similarity-based method, it incorporates neighbor information while extracting embeddings for the target node. It takes a simple heuristic that considers the relevance of target user to users who previously are rated to target item alongside the users who have co-rating relation with the target user. However, this method does not need to extract features manually; instead, it utilizes a deep neural network model to extract recommendations. So it benefits from neighbor information while it supports offline training and online recommendation. We modified the standard attention mechanism to fix some flaws in previous methods in which attention has been applied in GNNs. In existing architectures of GAT (Graph Attention Network), the attention mechanism pay more attention to neighbor nodes rather than the target node itself, which is counterintuitive and can lead to low-quality final representations. Another drawback is that all neighbors may receive the same attention score from GAT because they consider graph weights as binary. In some nodes like influencers in social networks, it causes the final representation to become an average of many nodes. Also, most studies apply a single GNN layer with GAT on first-order neighbors (J. Song et al., 2021; W. Song et al., 2019; Zhong et al., 2020), which for a user are item nodes. This method results in an embedding that has a huge difference from the target user node. Also, some methods ignore weighted feedbacks provided by users and use them as a target for the training phase (J. Song et al., 2021). In all of these methods, a one-hot encoding is used as an initial feature vector which is not informative. So in this paper, we try to cover these drawbacks. Our major contributions in this paper can be summarised as follows:

- Presenting a hybrid Graph Neural Network and Matrix Factorization method to learn the representation of nodes in the graph
- Improving the performance of standard attention mechanism through some modifications
- Introducing a novel model-based collaborative filtering framework that outperforms state of the art baselines.

The rest of the paper is organized as follows: in section 2, we review the background of GNNs and their application in GRSs. We define the problem and introduce our method in section 3. Section 4 presents the experimental evaluations of our proposed approach and discusses our observations. Finally, we conclude our work with the future direction in section 5.  graph-structured data

## 2. Background

GNNs are deep learning models which learn from nodes information and topological structure of the graph (Fan et al., 2019; Ying, You, et al., 2018a). They are first introduced in (Gori, Monfardini, & Scarselli, 2005) to generate a representation for each node using a simple neural network that aggregates information like labels, representation of nodes, and edges. Then, they have been applied in different domains such as node classification, node clustering, node recommendation, link prediction, graph classification, and graph visualization (Cai, Zheng, & Chang, 2018). GNNs usually use aggregator and updater modules to embed nodes. The aggregator module aggregates the feature information from the neighbors, while the updater utilizes the aggregated information to update the embedding of the target node. GNN must decide how much information should be propagated from a node to its neighbors and how to aggregate the information received from the neighbors of a node, with or without considering the weights of the corresponding edges among nodes (Gao et al., 2020). It is possible to use different methods in aggregator and updater modules. For example, the neighbor information is aggregated using max-pooling and mean-pooling in (Ying, He, et al., 2018), while in (Hamilton, Ying, & Leskovec, 2017), an LSTM aggregator has been used. In the updator part, it is possible to consider the aggregated information as a new representation for the target node

(Berg, Kipf, & Welling, 2017; He et al., 2020); or aggregate this information with the original representation of the target node to get the final embedding (Hamilton, Ying, & Leskovec, 2017; Hekmatfar, Haratizadeh, & Goliaei, 2021).

Graph Convolutional Neural Network (GCN) is the most popular and frequently used GNN method (Zhou et al., 2020). Spectral-GCNs represent graphs based on their Laplacian matrix. Calculating the Laplacian matrix has a large computational time. It is also impossible to transfer a trained model from a specific structure to another graph. On the other hand, spatial-GCNs have some challenges in handling different sizes of neighbors and defining weight matrices for nodes. The GCN techniques have been successfully applied in several kinds of researches, using spectral construction (Defferrard, Bresson, & Vandergheynst, 2016; Henaff, Bruna, & LeCun, 2015) and spatial construction (Hamilton, Ying, & Leskovec, 2017; Monti et al., 2017; Niepert, Ahmed, & Kutzkov, 2016; Ying, You, et al., 2018b). Graph Attention Network (GAT) is a spatial-GCN method that adopts an attention mechanism (Bahdanau, Cho, & Bengio, 2014) in the propagation step of the GNN. It calculates attention coefficients for each neighbor based on its similarity to the target node, which is derived from their corresponding feature vectors (Veličković et al., 2017). GAT is computationally efficient, not structure-dependent, and allows information from different neighbors to aggregate with different weights. The attention mechanism is applied in GNNs in a variety of ways. It could assign a different attention coefficient to neighbors with different hops, as in DAGCN (F. Chen et al., 2019) and ChebyGIN (Knyazev, Taylor, & Amer, 2019). In a heterogeneous graph, attention is mainly used to assign different scores for each type of node and edge (like HetSANN (Hong et al., 2020) and RGHAT (Zhao Zhang et al., 2020)) or different meta-paths (Xiao Wang et al., 2019). In spectral GNN, GAT extracts the importance of different frequency components from low frequencies (the eigenvectors associated with small eigenvalues) to high frequencies (the eigenvectors associated with large eigenvalues) (Chang et al., 2020).

## 3. Related Work

Graph-based recommender systems are one of the application domains in which GNNs have been applied successfully. GNNs let model-based recommenders to incorporate information of multi-hop connections between users and items, which is helpful in collaborative filtering. Farther neighbors may also have helpful information about the target user's possible interests (S. Wu et al., 2020). The general framework for applying GNNs in the recommendation domain is presented in Figure 1. First, the data must be modeled as a graph structure that the GNN later analyzes it to embed the important entities, including users and items. These embeddings are then used to predict a user's possible interest in each item, based on which the recommendation list is generated (Gao et al., 2020; S. Wu et al., 2020). In different studies, various graphs have been used to model the information, and the final recommendation has been extracted with the help of different algorithms. However, the main difference between the methods is in the way they represent users and products.

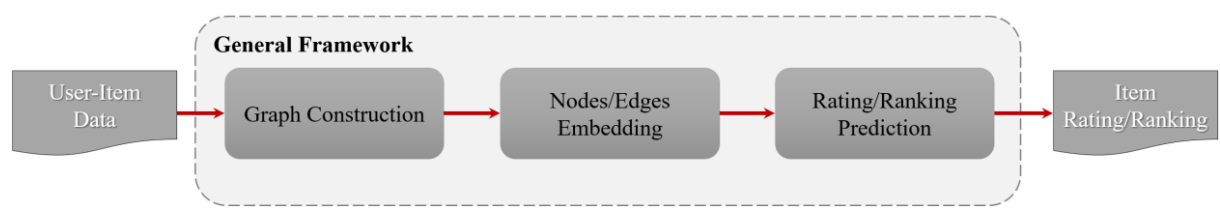

*Figure 1. GNNs structure in the recommender systems*

The first GNN proposed to embed nodes in the recommender graph was MGCNN (Monti, Bronstein, & Bresson, 2017). It incorporates side information along with user feedback in the recommendation extraction process. They modified the convolution operation in ChebNet

(Defferrard, Bresson, & Vandergheynst, 2016) to extract user and item embeddings simultaneously from user-user and item-item graphs which are constructed based on side information. The extracted embeddings are improved progressively via a diffusion process using an RNN network to predict unknown feedback correctly. GC-MC is the first method that uses Graph Auto Encoders (GAE) in the recommender systems (Berg, Kipf, & Welling, 2017). Unlike MGCNN, which concentrates on node embeddings, GC-MC provides unknown ratings directly through a single, noniterative GNN. It incorporates a graph convolutional encoder-decoder to predict the weight of the edges between the user and item nodes. A GCN layer combines nodes' feature vectors based on the weight of the edge between them to encode nodes. A bilinear decoder gets encoded vectors for user and item and reconstructs user feedback. The PinSage (Ying, He, et al., 2018) was proposed to handle embeddings extraction challenges through GCN in a large-scale dataset like Pinterest. Instead of using all k-hop neighbor nodes, PinSage samples top $T$ similar undirect neighbors based on random walk scores, and after transforming them, it applies a pooling function. It concatenates the resulting embedding with the node's initial embedding and passes it through another dense layer to get the final representation. In GraphRec (Fan et al., 2019), a user is represented by its neighbor users in the social network and the items the user has interacted with. They use mean-pooling to aggregate information from neighbors. The unknown feedback is predicted with an MLP, which gets target item representation based on neighbor users and target user representation based on its user and item neighbors. SR-GNN (S. Wu et al., 2019) is the first GNN for the next item recommendation in session-based systems. Wu et al. have developed a gated GNN to learn embeddings of items in a session sequence graph. A session's embedding is then constructed by aggregating the embeddings of all item nodes in that session. Due to the different priorities of items in a session, the weight of each item in aggregating phase is calculated using the attention mechanism. Finally, multiplying each item's embedding by the session's embedding results in the score of each item. In Multi-GCCF (Sun et al., 2019), GCN is applied on $k$-hop neighborhood information in the user-item graph. The resulted embeding is merged with the generated embeding of one-hop neighbourhood GCN on the user-user and item-item graphs. These graphs are constructed based on pairwise cosine similarities of rows or columns of the rating matrix. In NGCF high-order indirect connectivity of users and items are considered in a user-item bipartite graph. In addition to propagating high-order neighbors information hop by hop and generating embeddings in each hop, they represent nodes final embeding by aggregating embeding embeddings from different order (Xiang Wang et al., 2019). In IG-MC (M. Zhang & Chen, 2019), for each user-item interaction, an enclosing subgraph containing user and items' $k$-hop neighbors is extracted, and unlike Pinsage and GC-MC, which applies node-level GCN (they extract two embedding for user and item), IG-MC applies a graph-level GCN on enclosing user-item subgraph and extracts an embedding for subgraph which models the interactions and relevance between the nodes. They first extract embedding for nodes in the subgraph through R-GCN, then aggregate nodes to represent the subgraph. To evaluate the social influence of each user (Q. Wu et al., 2019) applies a graph attention network for the first time on the recommender system. This model considers exclusive weights for each neighbor in the social network. To learn user preference over user's points of interest (POI) (Zhong et al., 2020), applied a multi-head attention mechanism on the product of POI embeddings from GCN and user previous interactions. The PGRec (Hekmatfar, Haratizadeh, & Goliaei, 2021) predicts the weight of links between users and preference nodes in a user-preference-item graph with one layer spectral GCN combined with factorization methods. To determine the amount of influence of each neighborhood in the final representation of the target node, a weight is calculated for each indirect neighbor based on the number of the paths between the target node and each neighbor. The GCE method (Duran et al., 2021) has been proposed to use context information and apply GCN on the user-context-item graph. A user node is represented based on the propagated information of items he interacts, as well as the context related to the interaction between the user and the item. GACSE (J. Song et al., 2021) considers a static embedding which is normalized average of all nodes and a dynamic embedding which is weighted average of neighbors calculated by graph attention mechanism. They proposed a combination of Bayesian Personalized Ranking and node's similarity as a ranking-oriented loss function.

## 4. Method

In this section, we describe the detail of our proposed method. The core of our method is generating high-quality node embeddings for user and item nodes in a recommender system graph. To do so, we apply a GCN that incorporates an attention mechanism. The attention mechanism helps GCN to use important neighbor information on extracting each node's embeddings. GARec is model-based collaborative filtering, and each node's embedding reflects a combination of information from its similar nodes inside an information graph (like a content graph) and co-rating relations.

### 4.1. Problem Setup

An essential part of the data in a recommendation system represents interactions among a set of users $U$ ($|U|=n$) with a set of items $I$ ($|I|=m$). Both users and items could have some meta-data (or content) $C$, such as the age of users or the genre of movies. The interaction matrix R contains $r_{ui}$ as an interaction between user $u \in U$ and item $i \in I$, which represents explicit feedback like the user's given a rating to a movie or implicit feedback like watching a movie in a VOD system. To show such information in a heterogeneous graph $G$, the main entities like users and items are usually modeled as nodes while an edge between a user and an item represents an interaction, possibly using a weight label. Figure 2 presents a sample of such a graph for explicit rating data, while the goal is to predict the weights of edges marked with '?'.

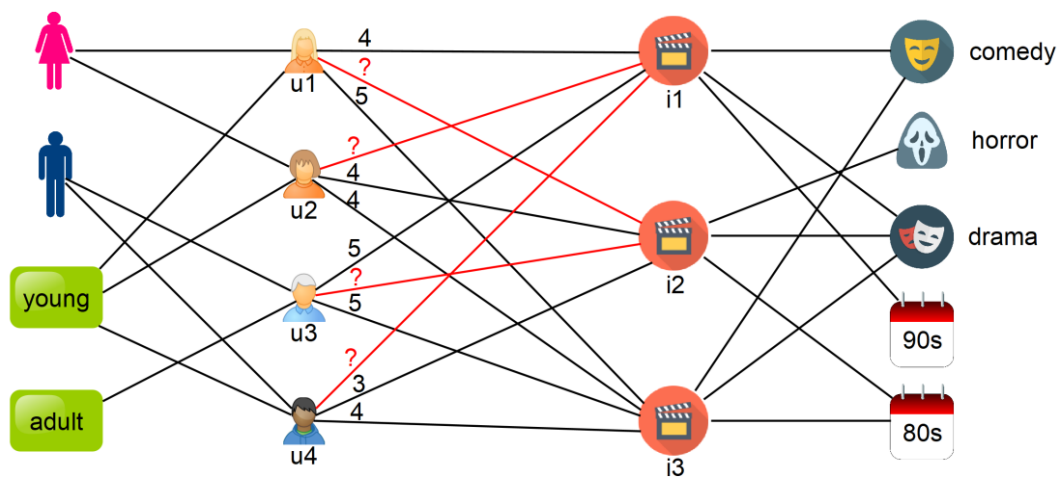

*Figure 2. A sample of heterogeneous recommender system graph*

In this graph, each user could connect to another user through instances of different meta-paths such as user-item-user or user-content-user. Likewise, the items could connect through item-user-item and item-content-item meta-paths.

### 4.2. Model Architecture

GARec is a model-based recommender system designed to predict the user $u$'s interest to an item $i$. It is an end-to-end model that gets user-item interaction matrix $R$ and the initial feature vector generated from $R$ ($F_U$ and $F_I$) and simultaneously learns to embed user and item nodes, extracting edge representation and predicting unknown ratings. First, a user-item bipartite graph ($G_{UI}$) is constructed from matrix $R$. The weight of the edges in $G_{UI}$ is the same as the user's feedback to the items and is specified by $W_{UI}$. GARec represents each edge between $u$ and $i$ based on the embedding of both $u$ and $i$ nodes. A spatial GCN model is responsible for calculating the representation nodes based on two types of indirect neighborhood for each node. So it is needed to identify informative neighbors and the amount of information propagated from it. Neighbors nodes could directly or indirectly connect to the target node. Some neighbors may not be relevant to the target node, and using their information may damage the representation of the nodes. So a Graph attention mechanism is used to identify similar neighbors and their relevancy called the attention coefficient. This coefficient determines how much each neighbor contributes to the embedding of the target node. The integration of the new embeddings for the user ($f_u'$) and item ($f_i'$) results in the final representation of the target edge. An MLP gets this representation and predicts the weight of the target edge. An overview of GARec is provided in Figure 3.

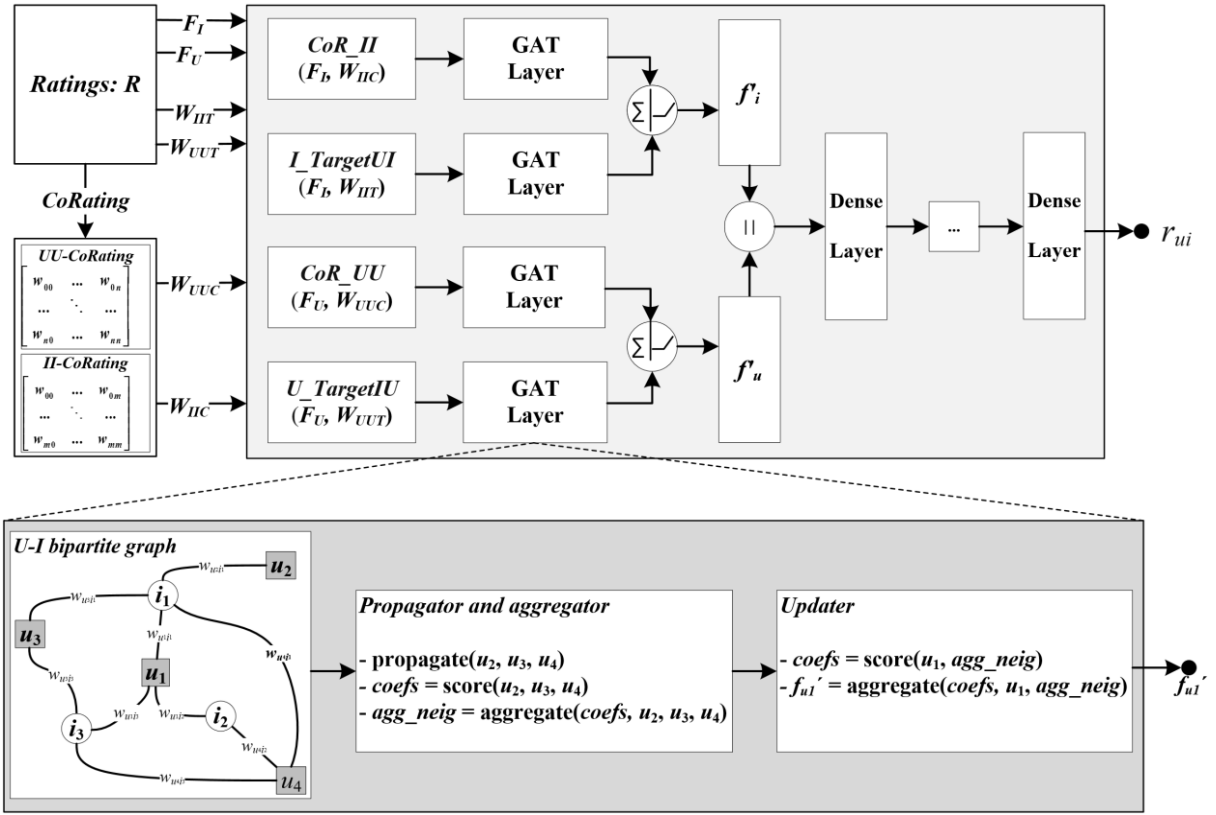

*Figure 3. Outline of proposed method GARec*

Applying the attention mechanism in graph neural networks has some challenges, and sometimes, it just increases the complexity of the model unnecessarily (Xu et al., 2018). Because of large attention to neighbor nodes, the generated embeddings for a node may have a significant distance from the nodes' initial embeddings. Such embeddings had low quality, and the network could not get high performance in prediction tasks with them. Also, if GAT generates similar attention coefficient for different nodes, the generated embedding for all nodes could be very similar to each other. In such settings, the network always predicts the same output for all inputs. To overcome these drawbacks we modified the graph attention layer in a way that we ensure the final embeddings contain node initial feature vector and its neighbors' information.

### 4.2.1. Neighbor selection

The proposed method is a spatial GCN-based model that uses the information of the neighbors of each node to represent it. In the GARec, each node is represented by nodes of the same type and with indirect connection (two-hop graph neighborhood) in the user-item bipartite graph. So to represent each user node, it is needed to select a number of neighbor users. Two types of neighborhoods are considered for a user; the first one is co-rated users (*UU-CoRating* in Figure 3). These users who can be effective in representing the target user are extracted from the user-item-user path. The weight of the indirect connection between user $y$ who has interacted with the same item $j$ as user $i$ is calculated as equation (1).

$$w_{uyC} = \sum_{i=1}^{I_u} r_{ui} \times r_{yi} \qquad (1)$$

In this equation, $I_u$ is a set of items that u has interacted with previously. $W_{UUC}$ in Figure 3 contains all these weights, and $W_{IIC}$ is calculated respectively for items that have got interaction from the same user.

The second types of neighborhood in embedding target users are users like $y$ who have provided feedback to the target item and recommending the target item $i$ will connect these nodes to the target node indirectly. We use the feedback $r_{yi}$ as $w_{uyT}$, the weight of the indirect relationship between user $i$ and $y$. $W_{UUT}$ in Figure 3 contains all these weights, and the same goes for items as $W_{IIT}$.

With the help of these weights, the proposed method for the target node could consider the relevance of co-rated neighbor users with users who interacted with the target item.

### 4.2.2. Initial Feature Vector

The proposed neural network model gets user and item information as input, and these inputs must be in the form of a vector. Some studies (J. Song et al., 2021; Q. Wu et al., 2019) use one-hot encoding as the initial feature vector; however, it projects users and items into a higher orthogonal dimensional the computationally space and time parameters dramatically increases. It ignores informative relationships existing among nodes. However, the matrix factorization methods could provide high-quality vectors for user and item nodes from the user-item interaction matrix. These methods analyze an interaction matrix $R$ that may contain implicit or explicit interactions between users and items and map both users and items to a joint latent factor space of size $d$. Each user $u$/item $i$ is associated with a vector $f_u$/$f_i$. Each element of these vectors measures the amount of a factor for a user/item (Koren, Bell, & Volinsky, 2009). There are several matrix factorization methods, including NMF (Luo et al., 2014), PMF (Ma et al., 2008), SVD++ (Koren, 2008), among which NMF is one of the most popular ones. It takes a non-negative matrix $R$ and produces two non-negative $F_U$:$n \times d$ and $F_I$:$m \times d$ ($d \ll n,m$) matrices in which $R \approx F_U.F_I^T = \tilde{R}$. Each row in $F_U$ is a feature vector $f_u \in \mathbb{R}^d$ for a user $u$, while each row in $F_I$ is a feature vectors $f_i \in \mathbb{R}^d$ defining item $i$ (Koren, Bell, & Volinsky, 2009). The non-negativity of feature values make them more comprehensible, and if the value of a certain latent feature is large in both user and item vectors, the dot product will be large too. Intuitively that means if an item has a feature that a user likes, then it has a chance to receive a higher rate from that user (M. Wu, 2007). Also NMF's time complexity is less than most of the other matrix factorization techniques (Dai, Li, & Xiang, 2018) which makes it a simple and efficient way to extract initial feature vectors required by the GCN algorithm, so we use it for generating those vectors.

### 4.2.3. Aggregator

In each GNN layer, information is propagated from each node to its neighbors, and each node aggregates received information using an aggregator. There are various methods to aggregate neighbor information, like taking their elementwise summation, mean, max, etc. (Battaglia et al., 2018). In this study, we use the weighted mean, while the attention coefficients are weights used in this operation.

First, the initial feature vectors of target user $u$ ($f_u$) and its neighbor $y \in N_u$ ($f_y$), are transferred to another space using learnable weight $W_U$. The resulted vectors are $q_u$ and $k_y$, which are called query and key vectors. The relevance of $y$ to $u$ is calculated as equation (2).

$$rel_{uy} = a_{uy} \times (q_u \cdot k_y) \qquad (2)$$

Which $a_{uy}$ is the weight between user and item from weight matrices generated in the previous section. To remove irrelevant neighbors and keep neighbors with $rel_{uy} > 0$ we apply a ReLu function on relevance score. In the following, the $rel_{uy}$ is used as a coefficient to calculate the embedding of the target node, and softmax could make the sum of all $rel_{uy}$ values equal to 1. How ever the relevance values have long tail distribution, which applying softmax causes $rel_{uy} \approx 0$ on thinner side and $rel_{uy} \approx 1$ on the thicker side. So to handle the long-tail problem, we consider $rel_{uy} \approx -\infty$ for $rel_{uy}$ less than mean of all $rel$s, then normalize. So the resulting attention coefficient will be as equation (3).

$$coef_{uy} = \text{softmax}\left(\frac{\text{ReLU}(rel_{uy})}{\sum_{y \in N(u)} rel_{uy}}\right) \qquad (2)$$

This coefficient specifies the influence of neighbors on the final embedding of a target node. By aggregating the information of the neighbors based on the calculated coefficients, the amount of information received from all neighbors in the target node $u$ will obtain as *agg_neii* according to equation (4).

$$f_{u-nei} = \sum_{y \in N_u} coefs_{uy} \times q_u \qquad (4)$$

In this equation, $f_{u\text{-}nei}$ is aggregated vector of $N_u$ neighbor.

### 4.2.4. Updater

An updater function in GNN is responsible for calculating the updated embedding for each node. In different studies, various methods have been used for updater. Some may use a neural network like MLP, CNN, or RNN as an updater (Abadal et al., 2020). In (J. Song et al., 2021), the aggregated vector of neighborhood information is considered as the new representation for the target node. In (Veličković et al., 2017) a nonlinear transformation function is applied on $f_{u\text{-}nei}$ to get the final representation. Such methods of updating have two major drawbacks. Both mentioned methods ignore target nodes themselves and consider neighborhood only. Also, in the denser graphs, lots of nodes have the same neighbors who are influential nodes (nodes related to popular users or items, such as the top 250 movies in history). Therefore the proposed updator in (J. Song et al., 2021; Veličković et al., 2017) will generate similar embeddings for a large number of nodes, and their value will be the average of popular nodes.

To prevent such problems, a method similar to the aggregator is used. In this method, a separate attention component evaluates the relationship between the aggregated information and the target node as equation (5).

$$rel_{inei} = (q_u \cdot f_{u-nei} w_{u-nei}) \qquad (5)$$

Also, a score for the node's original feature vector is calculated to determine its initial feature vectors importance as equation (6).

$$rel_{ii} = (q_u \cdot f_u w_u') \qquad (6)$$

By applying weight matrix $w_u'$ and $w_{u\text{-}nei}$ on node's original feature vector $f_u$ and neighbors' aggregated featur vector $f_{u\text{-}nei}$, we transfer them to another dimension. Then we calculate their relevance to the query vector. By transferring to another space, we control the relevance score of nodes. To get the final representation of the target node, we use equation (6) to aggregate neighbor information with the target node.

$$f'_u = \sigma(rel_{uu} \cdot f_u + rel_{u-nei} \cdot f_{u-nei}) \quad (1)$$

The result of equation 6 ($f'_u \in \mathbb{R}^d$) is the new embedding for node *i* that can be used in the next step by the rating prediction module.

### 4.2.5. Rating Prediction

To predict unknown ratings/rankings, different approaches have been used in various studies. Using MLP that takes user and item embedding and predicts unknown ratings/rankings is one of the simple ways that has been applied in lots of studies (Duran et al., 2021; Fan et al., 2019; X. Wang et al., 2020). So in the proposed method, at the final stage, the embeddings of users and items are given to an MLP to predict the rates. Here we use an end-to-end training technique, which is a popular way to train a model with multiple modules and one main objective, and usually leads to high performance in graph representation learning problems (Berg, Kipf, & Welling, 2017; Glasmachers, 2017). In this work, the objective of the whole model is to reduce the RMSE error of predicted weights for user-item edges. So, in the training phase, the whole model is trained to accurately predict the known weights of the edges between users and items, while the learning process adjusts the trainable parameters that exist in different modules of the model.

## 5. Experimental studies and Results

We designed some experiments to evaluate our proposed algorithm and demonstrate its effectiveness compared to state of the art baseline algorithm. We trained and evaluated the proposed model on a PC with the Ubunto 16.04 operating system and Cori5 Intel CPU, 16GB of Ram, and an NVIDIA 1050 Ti graphic card with 4GB memory. We implemented our proposed approach using Python version 3.5 and Tensorflow 2.2, and to boost the speed of executions, we run it on GPU.

### 5.1. Datasets and setup

To compare the performance of our proposed method with other baseline algorithms, we have chosen two public, well-known, and highly used datasets, MovieLens100K and MovieLens1M[1] (Harper & Konstan, 2016), which presents user's explicit feedbacks in the form of the 5-level rating system (1, 2, ..., 5). Each user has rated at least 20 movies in both datasets. Both datasets contain some extra data like age, gender, occupation, and zip code as users' side information, and genres, release date, and IMDB URL as movies' side information. We discretized user age and release date features, defining six classes for ages (as a teenager, young,…), and categorizing release dates into decades (the 40s, 50s,…). The details of the datasets are shown in Table 1.

*Table 1. The details of evaluated datasets*

| Datasets | Users | User side information | Movies | Movie side information | Ratings |
| --- | --- | --- | --- | --- | --- |
| **MovieLens100K** | 943 | Age, gender, zip code, occupation (21 occupations) | 1682 | Title, Movies genres (19 genres), release date, IMDb URL | 99992 |
| **MovieLens1M** | 6041 | Age range, gender, zip code, occupation (21 occupation) | 3779 | Title, Movies genres (18 genres), IMDB URL | 1000162 |

---

[1]They are available at GroupLens website https://grouplens.org/datasets/movielens/

We followed (Berg, Kipf, & Welling, 2017) experimental settings and conducted our experiments with 80/20 and 90/10 train/test splits and applied a 5-fold validation. We compared the performance of our model against other models with a widely-used evaluation metric for regression tasks which is RMSE (Root Mean Square Error).

### 5.2. Baselines
The proposed method is a model-based one, so we compared it with some of the other model-based methods: NMF(Luo et al., 2014), PMF (Ma et al., 2008), and SVD++ (Koren, 2008) methods are chosen as matrix factorization baselines while GC-MC (Berg, Kipf, & Welling, 2017), PinSage (Ying, He, et al., 2018), IGMC (M. Zhang & Chen, 2019), MCCF (X. Wang et al., 2020) and CF-NADE (Zheng et al., 2016) are selected as GNNs and AutoRec (Sedhain et al., 2015), Wide & Deep (Cheng et al., 2016) and DMF(Yi et al., 2019) are selected as neural network models.

### 5.3. Computational Complexity
The GARec algorithm consists of several independent components, so the computational complexity of the whole algorithm can be calculated as a combination of the computational complexity of each component.

The heterogeneous graph used in this algorithm consists of $n$ user nodes and $m$ item nodesEach user node is connected to the nodes of the item layer with a maximum of $m$. On the other hand, each item connects to users with a maximum number of $n$ edges. Therefore, in the densest case, the heterogeneous graph will have a total of $m+n$ nodes and $nm$ edges. However, in reality, this graph has a smaller number of edges due to low user feedback on items. Therefore, user-user, user-side information, and item-content subgraphs can be modeled with sparse proximity matrices. The number of user feedback on the item is limited, and $c$ can be considered the average number of user feedback, and the number of user-item subgraph edges can be considered $c \times m$. The two other matrices, user-user ($n \times n$) and item-item($m \times m$), which are defined based on the number of paths of length two in the user-item -user and item-user-item paths, each can be defined by the complexity $O(n \times n)$ and $O(m \times m)$ ($c_I$, $c_U << n$).

The proposed algorithm first generates the initial feature vectors of users and items using the NMF method with time complexity $O(n \times m)$ (Dai, Li, & Xiang, 2018).

Attention mechanism has a matrix multiplication operation, so due to the complexity of multiplication, applying this mechanism per user-item, user-user, and item-item graphs, in the worst case, will have $O(nm)$, $O(n^2)$, and $O(m^2)$ computational complexity (Ziwei Zhang, Cui, & Zhu, 2020). Also, the final prediction will be generated using an MLP whose complexity based on the number of training samples is $O(n \times m)$.

Therefore, the most expensive part of the algorithm, assuming $n>m$, has a time cost of $O(n^2)$. Table 2 compares the time complexity of the GARec algorithm against other algorithms.

*Table 2. The time complexity of the GARec algorithm compared to other algorithms*

| Algorithm | Computational Complexity | Description |
|---|---|---|
| **GARec** | $O(n^2)$ | |
| **NMF** | $O(mn)$ (Dai, Li, & Xiang, 2018) | |
| **SVD++** | $O(m^3)$ (S. Wang, Sun, & Li, 2020) | |
| **GC-MC** | $O(m)$ (Ziwei Zhang, Cui, & Zhu, 2020) | |
| **IG-MC** | $O(kmn)$ (M. Zhang & Chen, 2019) | *k is the number of subgraphs extracted for each node* |
| **PinSage** | $O((n+m) s^l)$ (Ziwei Zhang, Cui, & Zhu, 2020) | *l is the number of layers, s is the number of sampled nodes* |

## 5.4. Comparison Result

The performance of the algorithms in each dataset is presented in Table 3. The performance of the proposed method is reported over five independent experiments, but for the baseline algorithms, the best result has been reported.

*Table 3. Comparison between the proposed method and baseline algorithms in terms*

| Methods | MovieLens-100K | | MovieLens-1M | |
|---|---|---|---|---|
| | 80-20 | 90-10 | 80-20 | 90-10 |
| NMF | 0.963 | 0.958 | 0.917 | 0.914 |
| SVD++ | 0.946 | 0.913 | 0.855 | 0.847 |
| PMF | 0.939 | 0.924 | 0.868 | 0.883 |
| AutoRec | 0.943 | 0.934 | 0.889 | 0.844 |
| Wide & Deep | 0.948 | 0.937 | 0.891 | 0.869 |
| DMF | 0.892 | 0.889 | 0.836 | 0.828 |
| GC-MC | 0.905 | 0.891 | 0.847 | 0.832 |
| PinSage | 0.962 | 0.951 | 0.919 | 0.906 |
| IGMC | 0.927 | 0.905 | 0.860 | 0.857 |
| MCCF | 0.907 | 0.891 | 0.855 | 0.841 |
| CF-NADE | 0.922 | 0.911 | 0.848 | 0.829 |
| GARec | 0.880 | 0.872 | 0.811 | 0.803 |

## 5.5. Discussion

It can be observed in the reported results that all algorithms have lower RMSE when they have more data in the training phase, and it also applies to the proposed method. GNN-based methods perform better than factorization and traditional neural networks. According to experimental results, GARec improves other GNN methods' performance. In MovieLens-100K dataset with 80% of train data, GARec outperforms SVD++ as a top matrix factorization 0.7, AutoRec as a Neural Network model as 6%, and GC-MC as a graph neural network model 2% respectively. GNN-based recommenders benefit from both graphical modeling and neural network nonlinearity. So they can extract indirect relations between entities and aggregate information from those entities before mapping it to the target variable, and that is the reason for the superiority of GNNs over traditional neural networks. The existence of graphs and valuable information extracted from it seems to have made GARec perform better than methods without graphs like NMF, PMF, and SVD ++ as matrix analysis methods, or NADE, AutoRec, and Wide & Deep as neural network-based methods. The MLP part in GARec is similar to the prediction part in other GNN methods; however, the quality of embeddings is the key factor that improves overall algorithm performance. Modifying the attention mechanism in both aggregator and updator resulted in better representation which increased the performance of the algorithm compared to the competitors. Also, better initial vectors can improve final embeddings too. Unlike the Pinsage algorithm, which uses a Random Walker to identify important neighbors and prune unimportant ones, the GARec algorithm can identify important neighbors within the attention mechanism. Therefore, GARec uses a model to identify neighborhoods, which due to the advantages of model-based methods over random pacing (such as better performance in sparse data), this way of generating and selecting neighborhoods can be one of the reasons. GARec is superior to Pinsage.

## 6. Conclusion

The attention mechanism is a novel technique in the field of machine learning, especially when applied in the domain of graph-based analysis. In this paper, we proposed a graph-based recommendation algorithm, GARec, as a Graph Neural Network which applies spatial graph convolutional neural network with attention mechanism on a graphical model of recommender system data. It predicts the weights of the edges connecting users to items in a user-item bipartite graph. To do so, it extracts embeddings user and item nodes by aggregating information from their relevant neighbors. The main idea in this work is to find important neighbors for a user or item and pay more attention to information coming from those neighbors when generating embeddings. Attention mechanism could help to find out how much each neighbor should attend in a target node's embedding process. By modifying the attention mechanism and applying it in the updater module of GNN, we also ensure that the embedding process considers the target node's information as well. Experiments show that GARec improves previous methods' performance in terms of RMSE metrics in different datasets.

The underlying graph and the proposed recommendation method can be easily used for implicit feedback data too, but still, it can not handle certain kinds of information like sequential user-item interactions, as commonly exist in session-based data. One possible direction for future research would be to revise the model, including the graphical modeling and the graph analysis modules, to support the sequential information as well. Also, it is worth considering different definitions of the neighborhood in applying attention mechanisms, like considering nodes that are sampled in the random walk as neighbors.